**PAPER • OPEN ACCESS**

# Thermally enhanced photoluminescence for energy harvesting: from fundamentals to engineering optimization

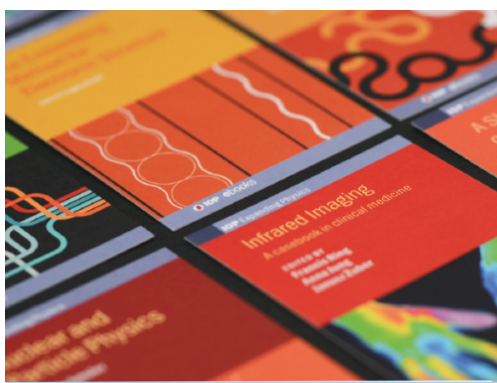

# Thermally enhanced photoluminescence for energy harvesting: from fundamentals to engineering optimization

**N Kruger**[1,4]**, M Kurtulik**[2]**, N Revivo**[3]**, A Manor**[2]**, T Sabapathy**[3] **and C Rotschild**[1,2,3,5]

[1] Grand Energy Program, Technion—Israel Institute of Technology, Haifa 32000, Israel
[2] Russell Berrie Nanotechnology Institute, Technion—Israel Institute of Technology, Haifa 32000, Israel
[3] Department of Mechanical Engineering, Technion—Israel Institute of Technology, Haifa 32000, Israel

E-mail: nkruger@tx.technion.ac.il

## Abstract

The radiance of thermal emission, as described by Planck's law, depends only on the emissivity and temperature of a body, and increases monotonically with the temperature rise at any emitted wavelength. Non-thermal radiation, such as photoluminescence (PL), is a fundamental light–matter interaction that conventionally involves the absorption of an energetic photon, thermalization, and the emission of a redshifted photon. Such a quantum process is governed by rate conservation, which is contingent on the quantum efficiency. In the past, the role of rate conservation for significant thermal excitation had not been studied. Recently, we presented the theory and an experimental demonstration that showed, in contrast to thermal emission, that the PL rate is conserved when the temperature increases while each photon is blueshifted. A further rise in temperature leads to an abrupt transition to thermal emission where the photon rate increases sharply. We also demonstrated how such thermally enhanced PL (TEPL) generates orders of magnitude more energetic photons than thermal emission at similar temperatures. These findings show that TEPL is an ideal optical heat pump that can harvest thermal losses in photovoltaics with a maximal theoretical efficiency of 70%, and practical concepts potentially reaching 45% efficiency. Here we move the TEPL concept onto the engineering level and present Cr:Nd:YAG as device grade PL material, absorbing solar radiation up to 1 $\mu$m wavelength and heated by thermalization of energetic photons. Its blueshifted emission, which can match GaAs cells, is 20% of the absorbed power. Based on a detailed balance simulation, such a material coupled with proper photonic management can reach 34% power conversion efficiency. These results raise confidence in the potential of TEPL becoming a disruptive technology in photovoltaics.

[4] Author to whom any correspondence should be addressed
[5] This article belongs to the special issue: Emerging Leaders, which features invited work from the best early-career researchers working within the scope of the Journal of Optics. Dr Carmel Rotschild was selected by the Editorial Board of the *Journal of Optics* as an Emerging Leader.

Abundance of high efficiency solar conversion systems is paramount to a fossil fuel-free society. For photovoltaics (PVs), the single-junction solar cell is efficiency limited, as most of the energy is lost to heat by thermalization of energetic and sub-PV bandgap photons. For maximal solar concentration, the conversion efficiency ceiling is 41%, as described by the Shockley–Queisser (SQ) limit [1]. Ideas utilizing this heat loss for single-junction cells have been suggested over the years, e.g., solar thermal photovoltaics (STPVs) [2], but none have demonstrated significant conversion efficiencies as of yet. In STPV, an intermediate absorber is heated by concentrated sunlight, and its thermal emission is coupled to a PV cell. The highest recorded STPV efficiency [3] is 3.2%. The challenge of high temperature requirement and strict demand for heat retention constitute the major obstacle faced by this technology [4, 5]. Recently, thermally enhanced photoluminescence (TEPL) solar converters were proposed as a means to harvest thermal losses in photovoltaics, doubling their efficiency [6]. This concept, much like STPV, consists of a thermally insulated absorber illuminated by concentrated sunlight; only now a PL material is chosen as the absorber, and its PL emission is coupled to a PV cell.

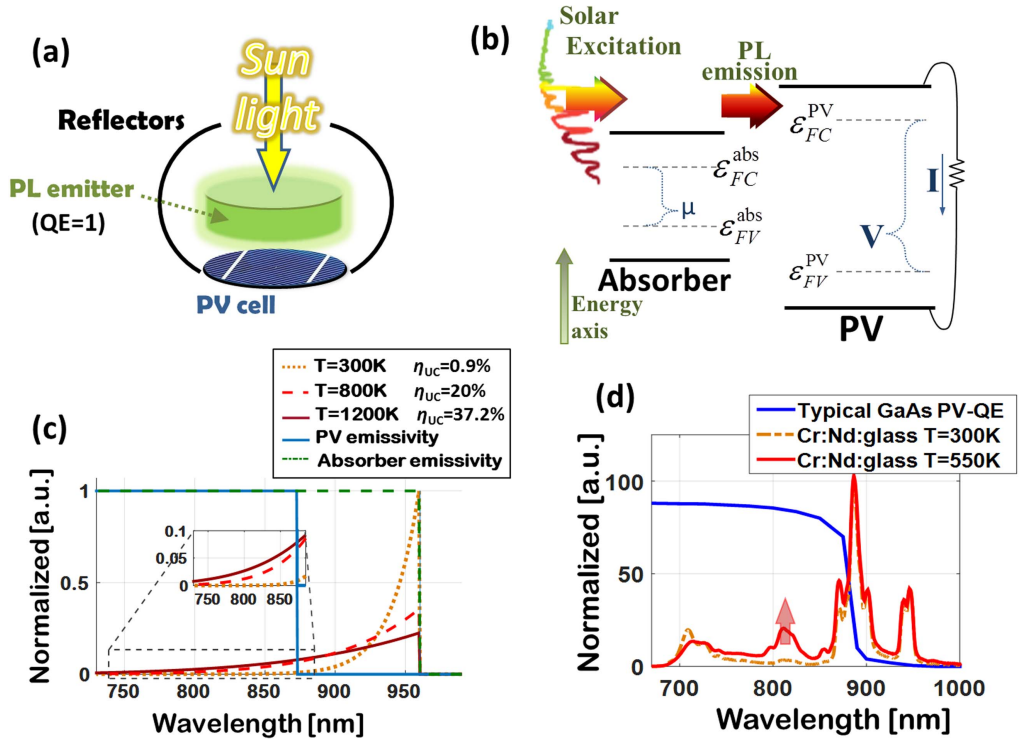

**Figure 1.** TEPL absorber-PV system. (a) Conceptual device design—an intermediate PL absorber placed to absorb sunlight and emit toward a PV cell, where all other emission angles are covered with reflectors for efficient photon recycling. (b) Energy diagram of such a device. (c) Step-function spectral response of an ideal PV with 1.42 eV bandgap (blue line) and relative emission of a step-function emissivity low bandgap 1.29 eV PL absorber at different temperatures (dotted orange line—300 K, dashed red line—800 K, solid dark-red line—1200 K). Up-conversion power efficiency above PV bandgap, $\eta_{UC}$, is given for each temperature. (d) Actual GaAs quantum efficiency (solid blue line) and measured Cr:Nd:YAG emission at room temperature (dashed orange line) and at 550 K (solid red line) by 532 nm laser excitation. Arrow indicates the TEPL effect.

In order to understand the difference between PL thermodynamics and thermal emission, we may address the following question: Let us say we heat a high quantum efficiency (QE) material to a temperature where its thermal emission, according to Planck's law, is 50 photons s$^{-1}$ at its band edge. Now we choose to excite this material with a 100 photon s$^{-1}$ excitation source. When asking what would be the emission rate under these circumstance, the answer is only 100 photons s$^{-1}$, and not, as one might think, 150 photons s$^{-1}$. The high temperature PL emission, compared to room temperature emission, is blueshifted so that each photon carries heat energy and acts as an optical heat-pump. Low bandgap PV cells absorb much of the solar flux, generating a high current, but at low voltage. By coupling the blueshifted emission of a low bandgap PL absorber to a high bandgap PV cell, we benefit from a high current and high voltage, potentially surpassing the SQ limit. The theoretical maximal efficiency for such a concept is 70% and a practical device is expected to work at efficiencies as high as 45% [7], exceeding the SQ limit of 32% and 38% for single junction solar cells under 1 and 1000 suns, respectively. These high efficiencies also require efficient photonic management, where non-absorbed photons are re-absorbed into the PL material. Note that a linear dependency exists between the pump and the emission rates, as defined by the QE. This situation contrasts with other frequency conversion processes, such as multi-photon absorption and high harmonic generation, where the nonlinear response competes with radiative and non-radiative decay processes. Figure 1 shows the device's main components (figure 1(a)) and energy diagram (figure 1(b)). Here the concentrated solar radiation is absorbed by a low bandgap PL material, placed inside a thermally insulated chamber. Absorption and thermalization of solar radiation elevates the temperature of the PL material, thereby blueshifting its emission, and matching it to a high bandgap solar cell.

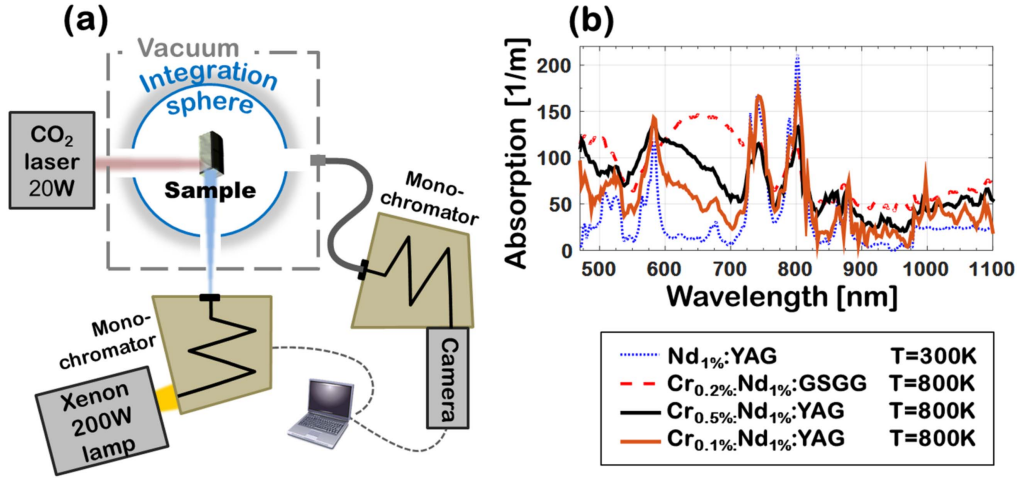

**Figure 2.** Experiment results (a) Calibrated solar simulation absorption and emission measurement setup. (b) Absorption of Nd:YAG, Cr:Nd:YAG and Cr:Nd:GSGG, at selected temperatures.

Thus far, TEPL was demonstrated only as a proof of concept with up-conversion efficiency of 2.5% at narrow band excitation [7], as the materials used were not optimized for maximal solar absorption or for retaining the high emission external quantum efficiency (EQE) at high temperatures. Here we study the performance of various absorber materials for TEPL under simulated solar excitation—specifically, $Nd^{3+}$ sensitized by $Cr^{3+}$ in YAG and GSGG crystal matrices. To better understand the criterion for a 'good' TEPL absorber, we examine its role as laid out in reference [7]. It must, as its name suggests, absorb the solar spectrum effectively, transforming this excitation into PL emission. This luminescence must remain efficient at high temperatures, emitting intensely at energies above the PV bandgap. Furthermore, the absorber is required to reabsorb the PL emission reflected back from the PV cell and walls to retain the exciton population, namely 'photon-recycling' (PR). This concept was previously used to increase PV efficiency [8] and is equivalent to angular restriction of emission.

In order to understand the different roles played by temperature, photon recycling, and quasi-chemical potential in the thermodynamics of PL, we first consider an ideal PL material, which has a step-function emissivity, with EQE of unity at all wavelengths. The emission is governed by thermodynamic properties and is a function of both the quasi-chemical potential and the temperature of the absorber, as described by the generalized Planck's law [9]:

$$R(\hbar\omega, T, \mu) = \varepsilon(\hbar\omega) \cdot \frac{(\hbar\omega)^2}{4\pi^2\hbar^3c^2} \frac{1}{e^{\frac{\hbar\omega-\mu}{k_BT}} - 1} \cong R_0 e^{\frac{\mu}{k_BT}}, \quad (1)$$

where $R$ is the emitted photon flux (photons per unit area per second). Here, $T$ is the temperature, $\varepsilon$ is the emissivity, $\hbar\omega$ is the photon energy, $k_B$ is Boltzmann's constant, and $\mu$ is the quasi-chemical potential (for describing the non-equilibrium radiation). In bandgap materials far from lasing, it is reasonable to assume that $\hbar\omega - \mu \gg k_BT$, leading to the simplification shown in equation (1) where the quasi-chemical potential describes the emission above the equilibrium thermal emission, $R_0$. Note that in practice, other properties also contribute to the description of the emission: internal energy transfer rates, internal QE and even the geometry of the absorber. We recently demonstrated that under elevated temperatures, the PL flux is conserved while the emission is blueshifted, meaning more photons are emitted at higher energy wavelengths (figure 1(c)—orange to dark red lines). This is true up to critical temperatures where the emission becomes thermal and the PL rate increases sharply [6]. For this reason, we choose a PV bandgap larger than the absorber bandgap, which enables us to extract a high voltage (figure 1(c)—blue line), while the sub-PV-bandgap emission is recycled via PR. This method allows us to benefit both from the high current of the low bandgap absorber and the high voltage of the solar cell, leading to device efficiencies potentially above the SQ limit. PR plays a critical role in converting the emission rate into electrical output above the SQ limit. For instance, a 1.29 eV PL emitter at 800 K has 20% of its emission rate with photon-energy higher than 1.42 eV (figure 1(c)—red dashed line). Furthermore, according to detailed balance calculation [7], if we take care to recycle even 85% of the lower energy photons we increase this value up to 34%. The 85% PR can be achieved by sub-bandgap reflectivity of 90%, and sample re-absorption of 20%.

In practice, optimizing an absorber material for a TEPL converter system has never been done. Organic [10] or semiconductor [11, 12] high-EQE emitting materials may be close to unity, however, their EQE is drastically reduced at temperatures needed for TEPL. Rare-earth (RE) doped matrices perform well under such conditions, with typical high luminescent efficiencies retained at high temperatures [6, 13–15]. Unfortunately, RE's absorption lines are typically narrow and do not absorb sunlight to the required degree. Sensitization of RE elements, which increases broadband absorption, is a well-known method used to increase efficiencies of flash lamp pumped RE lasers and solar powered lasers [16–21]. Specifically, neodymium (Nd$^{3+}$), frequently used for its high luminescent EQE, is promising for TEPL as its spectral emissivity contains lines below and above the bandgap of GaAs PV cells, allowing blueshifting of the high temperature PL emission above the bandgap (figure 1(d)).

Experimentally measuring a material's absorption and emission under solar excitation at different temperatures demonstrates its TEPL compatibility. As depicted in figure 2, we placed a sample of the absorber in the center of an integration sphere set inside a vacuum chamber. The sample was heated to high temperatures by a CO$_2$ laser [Synrad], while simultaneously excited by a selective excitation source of a monochromatically scanned xenon lamp [Newport] in the wavelength range of 450–1100 nm (figure 2(a)). The emission in the range of 650 nm to 1300 nm under each excited wavelength was measured both by an InGaAs and an Si camera [Andor]. The setup was calibrated by a QTH calibration lamp [Newport]. The measured emission was also used to determine the temperature of the absorber by the fluorescent intensity radiometry [6, 22] method, as the sample's emission was previously calibrated by a temperature-controlled furnace [MHI Inc.]. Later, the different spectrum measured for each excitation wavelength was superimposed at different weights so to simulate excitation by sunlight instead of a xenon lamp.

We measured several 1 wt% Nd$^{3+}$ doped YAG crystals, with additional Cr$^{3+}$ doping of 0, 0.1 and 0.5 wt%. The Cr$^{3+}$ acts as a sensitizer to Nd$^{3+}$ emission, allowing broadband absorption of the excitation spectrum. In addition, a Cr$_{0.2\%}$:Nd$_{1.0\%}$:GSGG crystal was also studied as an Nd$^{3+}$ host for lasers [20, 21]. All crystals were cut into 2 × 2 × 5 mm rectangular rods and excited on the 2 × 2 mm facet. The absorption and emission data were normalized for 63% absorption of the 580 nm Nd$^{3+}$ absorption line. For a non-sensitized Nd:YAG sample, this condition yielded 5% sunlight absorption (figure 2(b)—dotted blue). In comparison, the measured absorption of the Cr$_{0.5\%}$:Nd$_{1.0\%}$:YAG sample, was found to be 8% at room temperature, and 15% at 800 K, given that Cr$^{3+}$ absorption was drastically increased [14] (figure 2(b)—solid black line). For the GSGG sample, sunlight absorption was approximately 16%, and only slightly affected by temperature (figure 2(b)—dashed red line).

We measured the emission of the samples under simulated one-sun excitation. Figure 3 shows the emission at selected temperatures. These spectra are put into the TEPL context by modeling the expected power output of an ideal solar cell exposed to each emission. In the model, we chose a step-function emissivity solar cell with a bandgap of 1.43 eV, as in GaAs PVs. We extracted the PV power output for each measured emission by a detailed balance calculation [1]. We then defined the efficiency $\eta$—the ratio of PV power-out to solar power (1.5AM solar spectrum), and the internal-efficiency $\eta_{int}$—the ratio of PV power-out to absorber *absorbed* power. The latter indicates system potential given higher dopant concentrations and/or longer absorber lengths. We stress here that actual TEPL efficiency is potentially much greater, given the higher quasi-chemical potentials induced by solar concentration and PR schemes and reduced self-absorption by surface and geometrical optimization. Calculating the efficiency for non-sensitized Nd$^{3+}$ produced $\eta = 0.4\%$ at room temperature, and $\eta_{int} = 3.2\%$ when accounting for the absorption of this sample (figure 3(a)—dotted blue line). Specifically, for this sample, increased temperature led to parasitic absorption by the matrix; $\eta_{int}$ was normalized by room temperature absorption only. Raising the temperature up to 900 K led to a steady increase, up to $\eta = 2.4\%$ and $\eta_{int} = 20\%$ (figure 3(a)—solid blue). For Cr$^{3+}$ sensitized Nd$^{3+}$ in a GSGG matrix, results were typically low, no more than $\eta = 1.3\%$ and $\eta_{int} = 3.6\%$ at $T = 650$ K (figure 3(a)—solid red line). On the other hand, measuring the Cr:Nd:YAG samples showed far greater emission intensities. Specifically, the Cr$_{0.5\%}$:Nd$_{1\%}$:YAG sample showed room temperature efficiencies of $\eta = 2\%$ and $\eta_{int} = 10\%$, rising to $\eta = 4\%$ and $\eta_{int} = 15\%$ at 750 K, whereas the Cr$_{0.1\%}$:Nd$_{1\%}$:YAG also reached $\eta_{int} = 20\%$ at 750 K (figure 3(b)). These numbers demonstrate that longer samples, optimized for full absorption of solar radiation, will reach $\eta = 20\%$.

Close examination revealed that the main Nd$^{3+}$ emission lines (830, 890 and 930 nm) in Cr$^{3+}$ sensitized systems are prone to blueshifting when temperature increases, as described in previous work [7]. In contrast to un-sensitized Nd:YAG (figure 3(a)—blue lines), however, the 890 and 930 nm lines emission rates increased with temperature, while the Cr$^{3+}$ 700 nm emission line gradually declined. This elevated emission can be attributed to Cr$^{3+}$ absorption [14], and to the prolonged emission lifetime of Cr$^{3+}$ sensitized neodymium [23], when the temperature was increased. This behavior allows nearly doubling of the external efficiency at elevated temperature by Cr$^{3+}$ sensitization, while maintaining the internal efficiency of the Nd$^{3+}$ acceptor. Further temperature increase in the measured YAG single crystals, beyond 750 K, however, led to decreased emission and reduction in overall performance.

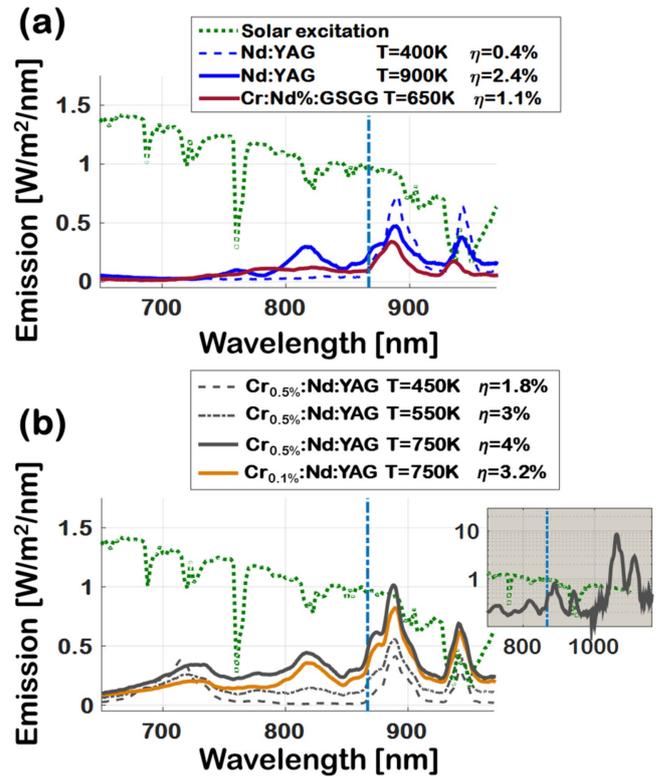

**Figure 3.** Measured emissions. Emission spectra at sub-optimal temperatures (dashed lines) and optimal temperatures (solid lines) under one-sun excitation (dotted green line) for (a) Nd$_{1\%}$:YAG, Cr$_{0.2\%}$:Nd$_{1\%}$:GSGG, (b) Cr$_{0.1\%}$:Nd$_{1\%}$:YAG, and Cr$_{0.5\%}$:Nd$_{1\%}$:YAG samples. The vertical blue line is the 1.43 eV ideal GaAs cut-off wavelength. Inset: log-scale emission of Cr$_{0.5\%}$:Nd$_{1\%}$:YAG sample including the 1 μm emission, typical of all Nd$^{3+}$ spectra.

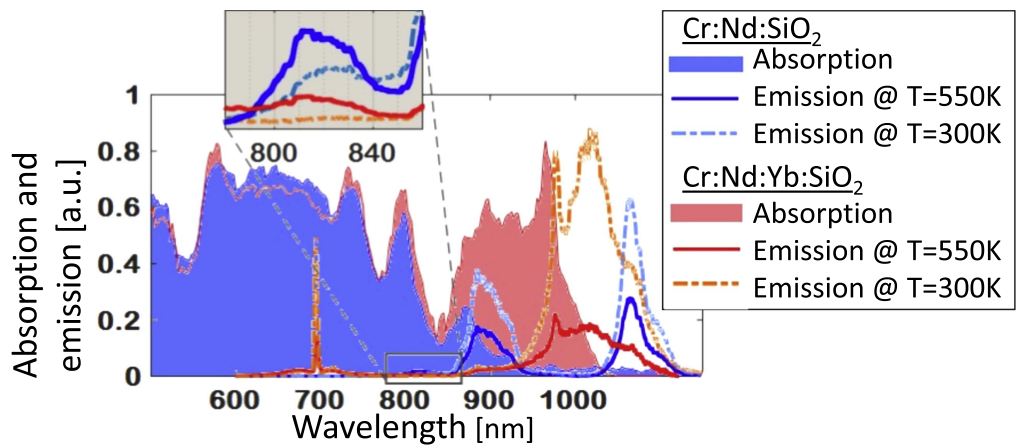

**Figure 4.** Absorption (filled area) and emission at room temperature (dotted lines) and 550 K (solid lines) of Cr:Nd (blue lines) and Cr:Nd:Yb (red lines) in glass. Inset: zoom-in on the emission of the 830 nm Nd$^{3+}$ emission line—presenting the expected blueshifting at elevated temperatures. Absorption was measured as with the YAG samples in figure 2(b). Emission, spectrally calibrated, is presented in arbitrary units and measured under 532 nm laser excitation. Heating to 550 K was done in a furnace while temperatures were measured by a thermocouple.

As noted above, the $\eta_{int}$ efficiency indicates the bottom threshold of any TEPL device, raising the option of effective enhancement of this value by PR schemes. One measure of the efficiency of any PR scheme is the self-absorption of sub-bandgap photons, with respect to the high bandgap PV. Typical 1.06 $\mu$m Nd$^{3+}$ emission lines are an order of magnitude brighter than the emission at the 700–870 nm band, above the PV bandgap (figure 3(b)—inset). In order to recycle this 1.06 $\mu$m wavelength emission, the self-absorption of the absorber must be balanced with the photon recycling quality, similar to balancing absorption and mirror losses in lasers. For example, given a self-absorption coefficient of $\alpha \approx 30 m^{-1}$ above a 1 $\mu$m wavelength (figure 2(b)) and a 2 mm wide sample, reabsorption of more than half of this emission would require mirror reflection greater than 94%. Increasing the absorption in this regime, given the same design parameters, will allow for higher recycling efficiencies. Moreover, we saw that adding acceptor dopants, such as ytterbium [6], may offer a significant advantage. Yb$^{3+}$ absorption, up to 980 nm, and emission, 0.98–1.05 $\mu$m (figure 4), will increase self-absorption of sub-bandgap emission, allowing for increased PR efficiencies. In addition, energy transfer to Yb$^{3+}$ reduces Nd$^{3+}$ emission of above bandgap photons, a process reversed at elevated temperatures [6]. Hence, maintaining high EQE at high temperatures will be vital in such a system.

Using the broadband laser pumping material Cr:Nd:YAG, enabled us to record efficiency results. Upgrading to such a state-of-the-art system, using a YAG ceramic matrix [14, 24, 25], as opposed to a single crystal structure, may increase this value even further due to higher EQE at high temperatures. Moreover, our absorber material in not restricted by thermal expansion properties or narrow line emission as laser materials are. We may, therefore, in future work consider adding additional donor dopants, such as cerium [26].

To conclude, we demonstrated that conventional Cr:Nd sensitization for TEPL purposes supports 20% internal efficiency and 4% external efficiency. Based on the detailed balance simulation, improving the external QE to above 20% and placing this material in a TEPL device, comprising more than 85% PR, will increase the conversion efficiency to above 34%. Further research in specific dopants tailored for TEPL together with refining the choice of transparent matrix will, potentially, lead to achieving the higher than 45% efficiencies expected from TEPL solar conversion technology.

## Acknowledgments

The research leading to these results was supported financially by the European Union's Seventh Framework Programme (H2020/2014-2020]) under grant agreement no. 638133-ERC-ThforPV. N Kruger acknowledges the support of the Nancy and Stephen Grand Technion Energy Program (GTEP). A. Manor acknowledges the support of the Adams Fellowship program. M Kurtulik acknowledges the support of the Russell Berrie Nanotechnology Institute (RBNI).

## ORCID iDs

C Rotschild  https://orcid.org/0000-0001-6400-9301

## References

[1] Shockley W and Queisser H J 1961 *J. Appl. Phys.* **32** 510
[2] Wurfel P and Ruppel W 1980 *IEEE Trans. Electron Devices* **27** 745–50
[3] Lenert A, Bierman D M, Nam Y, Chan W R, Celanović I, Soljačić M and Wang E N 2014 *Nat. Nanotechnol.* **9** 126–30
[4] Ferrari C, Melino F and Bosi M 2013 *Sol. Energy Mater. Sol. Cells* **113** 20–5
[5] Nam Y, Yeng Y X, Lenert A, Bermel P, Celanovic I, Soljačić M and Wang E N 2014 *Sol. Energy Mater. Sol. Cells* **122** 287–96
[6] Manor A, Martin L and Rotschild C 2015 *Optica* **2** 585–8
[7] Manor A, Kruger N, Sabapathy T and Rotschild C 2016 *Nat. Commun.* **7** 13167
[8] Braun A, Katz E A, Feuermann D, Kayes B M and Gordon J M 2013 *Energy Environ. Sci.* **6** 1499
[9] Wurfel P 1982 *J. Phys. C: Solid State Phys.* **15** 3967–85
[10] Nakanotani H, Higuchi T, Furukawa T, Masui K, Morimoto K, Numata M, Tanaka H, Sagara Y, Yasuda T and Adachi C 2014 *Nat. Commun.* **5** 4016
[11] Schnitzer I, Yablonovitch E, Caneau C and Gmitter T 1993 *Appl. Phys. Lett.* **62** 131–3
[12] Bender D A, Cederberg J G, Wang C and Sheik-Bahae M 2013 *Appl. Phys. Lett.* **102** 252102
[13] Gouveia E A, de Araujo M T and Gouveia-Neto A S 2001 *Braz. J. Phys.* **31** 89–101
[14] Honda Y, Motokoshi S, Jitsuno T, Miyanaga N, Fujioka K, Nakatsuka M and Yoshida M 2014 *J. Lumin.* **148** 342–6
[15] Torsello G, Lomascolo M, Licciulli A, Diso D, Tundo S and Mazzer M 2004 *Nat. Mater.* **3** 632–7
[16] Liang D, Almeida J and Guillot E 2013 *Appl. Phys.* B **111** 305–11
[17] Liang D, Almeida J and Garcia D 2013 *Proceedings Volume 8785, 8th Iberoamerican Optics Meeting and 11th Latin American Meeting on Optics, Lasers, and Applications* 87859Y
[18] Lupei V, Pavel N and Lupei A 2014 *Laser Phys.* **24** 045801
[19] Reusswig P D, Nechayev S, Scherer J M, Hwang G W, Bawendi M G, Baldo M A and Rotschild C 2015 *Sci. Rep.* **5** 14758
[20] Thompson G A, Krupkin V, Yogev A and Oron M B 1992 *Opt. Eng.* **31** 2644–6
[21] Reed E 1985 *IEEE J. Quantum Electron.* **21** 1625–9
[22] Berthou H and Jörgensen C K 1990 *Opt. Lett.* **15** 1100–2
[23] Saiki T, Nakatsuka M, Fujioka K, Motokoshi S, Imasaki K and Iida Y 2013 *Opt. Photonics Lett.* **6** 1350003
[24] Sokol M, Kalabukhov S, Kasiyan V, Dariel M P and Frage N 2016 *J. Am. Ceram. Soc.* **99** 802–7
[25] Saiki T, Fujiwara N, Matsuoka N, Nakatuka M, Fujioka K and Iiida Y 2017 *Opt. Commun.* **387** 316–21
[26] Fujioka K, Saiki T, Motokoshi S, Fujimoto Y, Fujita H and Nakatsuka M 2013 *J. Lumin.* **143** 10–3